\documentclass{iau_FM}
\usepackage{graphicx}
\usepackage{natbib}
\usepackage{mathtools}

\title[Hot Star Winds] 
{Hot Star Winds}

\author[A.A.C. Sander]   
{Andreas A.C. Sander$^1$}

\affiliation{$^1$Zentrum für Astronomie der Universität Heidelberg, Astronomisches Rechen-Institut, Mönchhofstr. 12-14, 69120 Heidelberg, Germany \\ email: {\tt andreas.sander@uni-heidelberg.de}} 

\pubyear{2024}
\jname{Astronomy in Focus, Focus Meeting 4} 
\editors{Hanindyo Kuncarayakti, Keiichi Maeda, eds.}
\begin{document}

\maketitle

\begin{abstract}
The properties, impact, and fate of hot stars cannot be understood without considering their winds. Revealed to be an almost ubiquitous phenomenon in the regime of massive stars, the winds of hot stars arise from a complex physical mechanism that still provides a major challenge for our understanding of massive stars. Different flavours of hot stars vary significantly in their winds with current evolution models still having problems to connect the zoo of observed phenomena. Moreover, the driving of hot star winds is inherently connected to the opacities arising from spectral line transitions, making the properties and strength of the winds strongly dependent on metallicity and changing them over cosmic time.

In these proceedings, the current status in our understanding of hot star winds is briefly reviewed and recent progress in our perception of hot star winds and the consequences for the evolution of massive stars are presented. A particular emphasis is given on current efforts with hydrodynamically-consistent atmosphere models towards a better description of radiatively-driven mass loss of Wolf-Rayet stars with remaining hydrogen envelopes.


\end{abstract}

\firstsection 
\section{Introduction: Hot stars and their radiation-driven winds}

\begin{figure}[htb]
\begin{center}
   \includegraphics[angle=0,width=\textwidth]{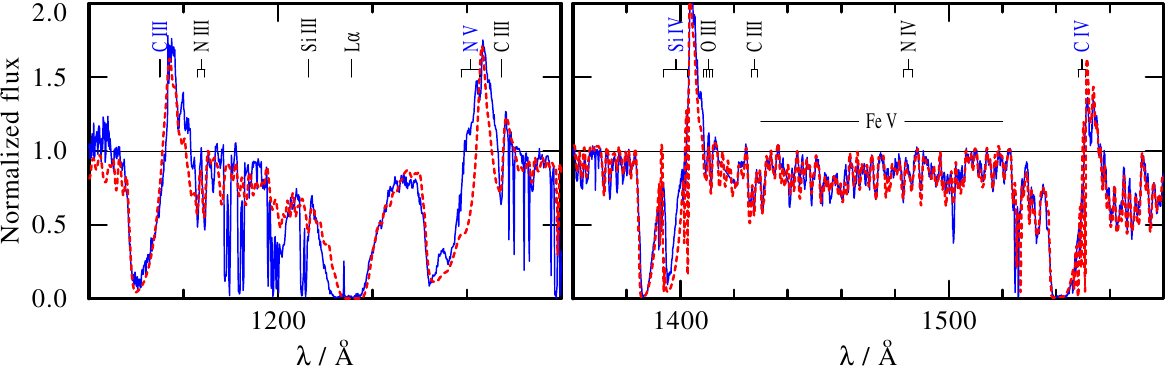}
     \caption{UV spectrum of the O9 supergiant Sk\,-66$^{\circ}$ 171 overplotted with the best-fit PoWR model. The characteristic P\,Cygni profiles (highlighted by blue idents) are important diagnostics for the stellar wind properties. The depicted analysis was performed within the ``XShooting ULLYSES'' collaboration \citep{Vink+2023} as part of the \citet{Sander+2024} study.}
   \label{fig:specana}
\end{center}
\end{figure}

Traditionally associated with a minimum initial mass of around eight to ten solar masses, massive stars are defined as stars that are intrinsically able to reach all nuclear burning stages in their interior and eventually undergo core collapse. With their luminosities already being high during their main sequence life, the upper part of the Hertzsprung–Russell diagram is mainly populated by massive stars. These stars spend the bulk of their lifetime -- in some cases even all of it -- as hot stars, characterized by $T_\mathrm{eff} > 10\,000\,$K. In this regime, associated with the spectral types B, O, and Wolf-Rayet (WR), the flux maximum of the stars is located in the ultraviolet (UV) regime, making the stars powerful sources of ionizing flux. Spectroscopy has revealed that this radiation is also key to understand the powerful stellar winds inherent to this regime (cf.\ Fig.\,\ref{fig:specana}). 

To sufficiently launch a wind, stars need to be close to the Eddington limit where gravity equals radiation pressure. For the classic limit, described by $\Gamma_\mathrm{e} = 1$ with
\begin{equation}
  \label{eq:Game}
  \Gamma_\mathrm{e} := \frac{\sigma_\mathrm{e}}{4\pi c G} \frac{L}{M}\mathrm{,}
\end{equation}
the radiation pressure only accounts for the opacity arising from Thomson scattering of free electrons. As evident from Eq.\,\eqref{eq:Game} proximity to $\Gamma_\mathrm{e} = 1$ largely depends on the ratio between luminosity $L$ and mass $M$. However, to overcome gravity, also the additional, depth-dependent opacities from spectral lines -- and sometimes also bound-free and free-free opacities -- have to be taken into account. Detailed models reveal a complex impact of the different elements and ions on the radiative force varying for different parameter regimes \citep[see, e.g.,][]{Sander2023}. Yet, iron opacities usually play the decisive role at wind onset and thus the resulting mass-loss rates $\dot{M}$ scale with the Fe abundance. In $\dot{M}$-recipes, the initial metallicity $Z$ is commonly taken as a proxy, but if $Z$ itself is estimated, e.g., from the gas-phase oxygen abundance, there can be severe errors if the host environment has a non-solar O/Fe ratio \citep[e.g., in IC\,1613 as discussed in][]{Bouret+2015}. 

As $L/M$ is intrinsically higher for more massive stars, radiation-driven winds are much weaker in most lower mass stars. However, there are notable exceptions as $L/M$ increases in later evolutionary stages and can be considerably boosted if the outer envelope of a star is removed, for example due to binary interaction. A subset of hot low-mass stars therefore shows considerable wind signatures, e.g., hot subdwarfs, and some hot objects in the low-mass regime even have WR-type spectra, indicating very strong winds.

\section{Wind regimes and mass-loss descriptions}

In general, one can distinguish between two kinds of radiation-driven winds in hot stars: \textit{OB-type winds} are optically thin at most wavelengths, in particular already in the launching region of the wind. In contrast, \textit{WR-type winds} are optically thick out to several stellar radii and arise only from stars closer to the Eddington limit. 
The spectral transition between O and WR is, however, not identical to the switch in the wind regimes. Instead, there is a transition regime where emission lines start to get prominent in the spectrum, while the winds stay mainly optically thin.

A widely applicable method to determine empirical wind parameters and in particular the mass-loss rate $\dot{M}$ is quantitative spectroscopy, where observed spectra are compared to synthetic spectra from detailed model atmosphere codes.
Recently, a comparison between the different codes and analysis methods has been performed by \citet{Sander+2024}. Finding generally good agreement, the study also reveals a method-dependent scatter of up to $0.4\,$dex in the derived $\dot{M}$, mainly due to differences in the reddening and clumping treatment. Due to their P\,Cygni-imprint (cf.\ Fig.\,\ref{fig:specana}), the terminal velocities $v_\infty$ are considered easier to measure. For OB stars, a clear correlation of $v_\infty$ with $T_\mathrm{eff}$ has been found \citep[e.g.,][]{Hawcroft+2024}, but with notable exceptions towards lower $v_\infty$, for example in case of evolved stars with more dense winds (see left panel of Fig.\,\ref{fig:envelope-mdot}).

The current theoretical state of the art for radiation-driven winds rests on a range of methods: For OB-star winds, \citet{Castor+1975} developed the (semi-)analytical ``CAK theory'' using a \textit{force multiplier} $\mathcal{M}$ to describe the ratio between radiative acceleration from spectral lines and Thomson scattering. Later, this concept was significantly extended \citep[e.g.,][]{FriendAbbott1986,Pauldrach+1986,OwockiPuls1999} and provides a fast, though approximate, way to prescribe the radiative acceleration. In parallel, Monte Carlo calculations established a second calculation pillar \citep[e.g.,][]{AbbottLucy1985,deKoter+1997} and provided the fundament for the widely-used OB-star mass-loss recipe from \citet{Vink+2001} as well as later follow-up efforts and extensions \citep[e.g.,][]{Vink+2011,Vink2017,VinkSander2021}. Growing computational capabilities more recently enabled a third approach to become feasible: model atmospheres combining a detailed comoving-frame radiative transfer with a locally-consistent solution of the stationary hydrodynamics. This technique has been utilized to establish new mass-loss predictions for OB-type winds \citep[e.g.,][]{Krticka+2020,Krticka+2021,Bjoerklund+2023} and in particular enabled to make progress in understanding the winds of WR stars \citep[e.g.,][]{GraefenerHamann2005,Sander+2020,SanderVink2020,Sander+2023}.

\section{Open questions and uncharted territory}

While the recent process in techniques have enabled new insights and $\dot{M}$-recipes, many open questions remain. Observations as well as new multi-dimensional radiation-hydrodynamic models of massive stars \citep[e.g.,][]{Moens+2022,Debnath+2024} seriously challenge current 1D treatments of important aspects shaping our ``empirical'' mass-loss measurements, e.g., regarding the description of wind inhomogeneities or the consequences of radiation-driven turbulence in OB photospheres with higher $\Gamma_\mathrm{e}$. 

\begin{figure}[tb]
\begin{center}
   \includegraphics[angle=0,width=0.41\textwidth]{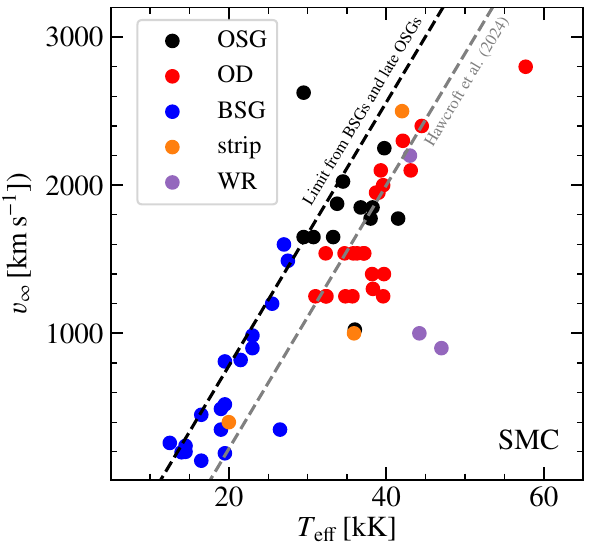} \hfill
   \includegraphics[angle=0,width=0.575\textwidth]{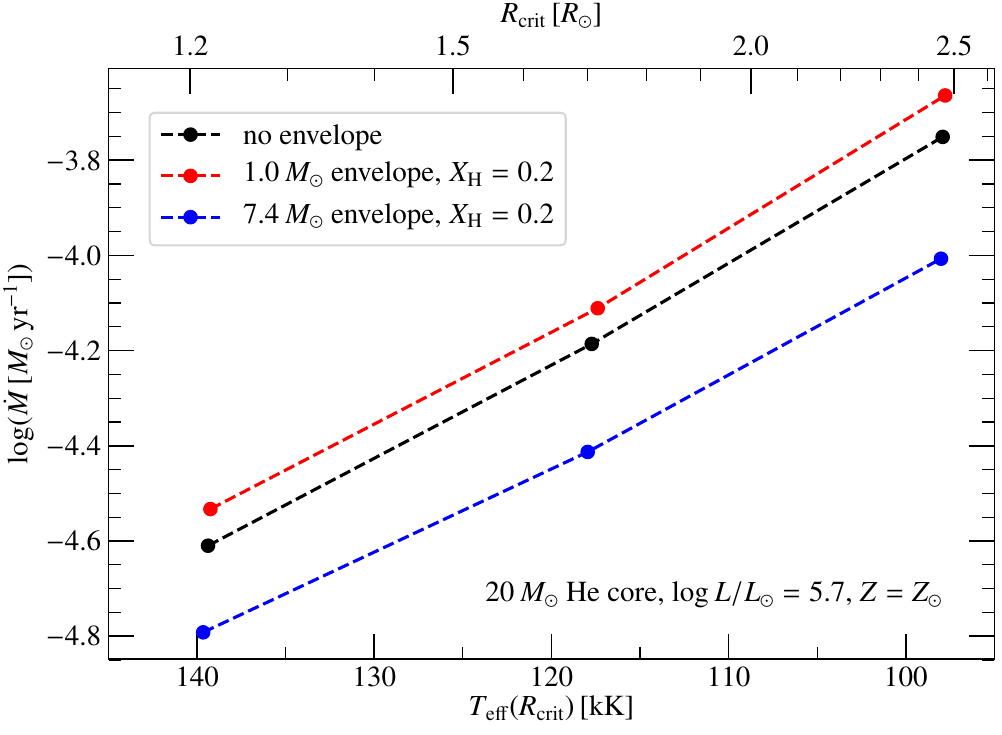}
     \caption{Left panel: Wind velocity versus $T_\mathrm{eff}$ of different types of hot stars in the SMC. The data are taken from \citet{Bouret+2013}, \citet{Hainich+2015}, \citet{Bernini-Peron+2024}, \citet{Ramachandran+2024}, and Backs et al. (submitted).
		  Right panel: Mass-loss predictions from consistent PoWR$^\textsc{hd}$ models assuming a $20\,M_\odot$ core-He burning star with different envelopes.}
   \label{fig:envelope-mdot}
\end{center}
\end{figure}

Beside these fundamental issues, there are also a lot of regimes that have hardly been studied, such as the sub-SMC metallicity regime, where many supernovae (SNe) occur.
Another highly uncharted territory is the mass loss of stars outside of the typical long-living stages of single-star evolution. High $L/M$ ratios naturally occuring in evolved stages can give rise to strong stellar winds, even in the low-mass regime, as illustrated by the discovery of the putative white-dwarf merger Pa30 \citep{Gvaramadze+2019,Lykou+2023}. Binary interaction can further produce objects with high $L/M$, thus boosting $\dot{M}$. A large uncertainty are in particular the mass-loss rates of massive stars with partially-stripped envelopes. Below the WR regime, many such objects are predicted, but few have been revealed \citep[e.g.,][]{Pauli+2022,Ramachandran+2023,Ramachandran+2024,Villasenor+2023}. Even for WR stars, the effect of a hydrogen envelope on top of a He-burning core is non-trivial (cf.\ Fig.\,\ref{fig:envelope-mdot}). If the envelope mass is very small, the $H$-presence leads to an increase in $\dot{M}$ as H atoms provide more free electrons than He atoms. Yet, any additional mass increases the gravitational force. For increasing envelope mass, the net effect on $\dot{M}$ would thus quickly become negative, but only when ignoring the structural response. A non-negligible envelope usually leads to an expansion of the outer layers, again lowering the gravitational force and thus yielding an increased $\dot{M}$ \citep[see also][]{Sander+2023}. In Fig\,\ref{fig:envelope-mdot}, this is quantified: The leftmost black data point marks the fiducial model for a $20\,M_\odot$ fully stripped core-He burning star. The blue data shows $\dot{M}$ if there is an additional envelope of $7.4\,M_\odot$. Even a modest structural expansion is sufficient to yield a positive change in $\dot{M}$, despite an increase of the total stellar mass by $>30$\%. 

Currently, none of these effects are properly accounted for in mass-loss descriptions for Wolf-Rayet stars. Detailed efforts coupling structure and atmosphere models will be necessary to synthesize a realistic description for partially-stripped massive stars. In the connection with SNe, further challenges exist. Both multiplicity and stellar winds, which are often interconnected rather than separate pathways, will be crucial to understand which types of SNe are associated with which types of massive stars. In particular, dedicated $\dot{M}$ descriptions for the pre-SN stages beyond core-He burning are so far missing.

\def\apj{{ApJ}}    
\def\nat{{Nature}}    
\def\jgr{{JGR}}    
\def\apjl{{ApJ Letters}}    
\def\aap{{A\&A}}   
\def\aaps{{A\&A Supplement}}  
\def\aapr{{The Astronomy and Astrophysics Review}} 
\def\araa{{Annu. Rev. Astron. Astrophys.}}
\def\mnras{{MNRAS}}
\def\nat{{Nature}}
\def\aj{{AJ}}
\def\pasa{{PASA}}
\def\pasp{{PASP}}
\def\ssr{{Space Science Reviews}}
\def\sovast{{Soviet Astronomy}}
\let\mnrasl=\mnras


\begin{thebibliography}{}

\bibitem[Abbott \& Lucy(1985)]{AbbottLucy1985}
{Abbott, D.~C. \& Lucy, L.~B.} 1985,
\textit{\apj}, 288, 679

\bibitem[Bernini-Peron et al.(2024)]{Bernini-Peron+2024} 
{Bernini-Peron, M., Sander, A.~A.~C., Ramachandran, V., et al.} 2024, 
\textit{\aap}, 
arXiv:2407.14216

\bibitem[Bj{\"o}rklund et al.(2023)]{Bjoerklund+2023} 
{Bj{\"o}rklund, R., Sundqvist, J.~O., Singh, S.~M., et al.} 2023, 
\textit{\aap}, 676, A109

\bibitem[Bouret et al.(2013)]{Bouret+2013} 
{Bouret, J.-C., Lanz, T., Martins, F., et al.} 2013, 
\textit{\aap}, 555, A1

\bibitem[Bouret et al.(2015)]{Bouret+2015} 
{Bouret, J.-C., Lanz, T., Hillier, D.~J., et al.} 2015, 
\textit{\mnras}, 449, 1545

\bibitem[Castor et al.(1975)]{Castor+1975}
{Castor, J.~I., Abbott, D.~C., \& Klein, R.~I.} 1975,
\textit{\apj}, 195, 157

\bibitem[Debnath et al.(2024)]{Debnath+2024}
{Debnath, D., Sundqvist, J.~O., Moens, N., et al.} 2024, 
\textit{\aap}, 684, A177

\bibitem[de Koter et al.(1997)]{deKoter+1997}
{de Koter, A., Heap, S.~R., \& Hubeny, I.} 1997, 
\textit{\apj}, 477, 792

\bibitem[Friend \& Abbott(1986)]{FriendAbbott1986} 
{Friend, D.~B. \& Abbott, D.~C.} 1986, 
\textit{\apj}, 311, 701

\bibitem[Moens et al.(2022)]{Moens+2022} 
{Moens, N., Poniatowski, L.~G., Hennicker, L., et al.} 2022, 
\textit{\aap}, 665, A42

\bibitem[Gr{\"a}fener \& Hamann(2005)]{GraefenerHamann2005} 
{Gr{\"a}fener, G. \& Hamann, W.-R.} 2005, 
\textit{\aap}, 432, 633

\bibitem[Gvaramadze et al.(2019)]{Gvaramadze+2019} 
{Gvaramadze, V.~V., Gr{\"a}fener, G., Langer, N., et al.} 2019, 
\textit{\nat}, 569, 684

\bibitem[Hainich et al.(2015)]{Hainich+2015} 
{Hainich, R., Pasemann, D., Todt, H., et al.} 2015, 
\textit{\aap}, 581, A21

\bibitem[Hawcroft et al.(2024)]{Hawcroft+2024} 
{Hawcroft, C., Sana, H., Mahy, L., et al.} 2024, 
\textit{\aap}, 688, A105

\bibitem[Krti{\v{c}}ka et al.(2020)]{Krticka+2020} 
{Krti{\v{c}}ka, J., Kub{\'a}t, J., \& Krti{\v{c}}kov{\'a}, I.} 2020, 
\textit{\aap}, 635, A173

\bibitem[Krti{\v{c}}ka et al.(2021)]{Krticka+2021} 
{Krti{\v{c}}ka, J., Kub{\'a}t, J., \& Krti{\v{c}}kov{\'a}, I.} 2021, 
\textit{\aap}, 647, A28

\bibitem[Lykou et al.(2023)]{Lykou+2023} 
{Lykou, F., Parker, Q.~A., Ritter, A., et al.} 2023, 
\textit{\apj}, 944, 120

\bibitem[Owocki \& Puls(1999)]{OwockiPuls1999} 
{Owocki, S.~P. \& Puls, J.} 1999, 
\textit{\apj}, 510, 355

\bibitem[Pauldrach et al.(1986)]{Pauldrach+1986} 
{Pauldrach, A., Puls, J., \& Kudritzki, R.~P.\ 1986}
\textit{\aap}, 164, 86

\bibitem[Pauli et al.(2022)]{Pauli+2022} 
{Pauli, D., Oskinova, L.~M., Hamann, W.-R., et al.} 2022, 
\textit{\aap}, 659, A9

\bibitem[Ramachandran et al.(2023)]{Ramachandran+2023} 
{Ramachandran, V., Klencki, J., Sander, A.~A.~C., et al.} 2023, 
\textit{\aap}, 674, L12

\bibitem[Ramachandran et al.(2024)]{Ramachandran+2024} 
{Ramachandran, V., Sander, A.~A.~C., Pauli, D., et al.} 2024, 
\textit{\aap}, accepted, arXiv:2406.17678

\bibitem[Sander et al.(2017)]{Sander+2017} 
{Sander, A.~A.~C., Hamann, W.-R., Todt, H., et al.} 2017, 
\textit{\aap}, 603, A86

\bibitem[Sander et al.(2020)]{Sander+2020} 
{Sander, A.~A.~C., Vink, J.~S., \& Hamann, W.-R.} 2020, 
\textit{\mnras}, 491, 4406

\bibitem[Sander \& Vink(2020)]{SanderVink2020} 
{Sander, A.~A.~C. \& Vink, J.~S.} 2020, 
\textit{\mnras}, 499, 873

\bibitem[Sander(2023)]{Sander2023} 
{Sander, A.~A.~C.} 2023, 
in IAU Symp.\ 370, \textit{Winds of Stars and Exoplanets}, 130

\bibitem[Sander et al.(2023)]{Sander+2023} 
{Sander, A.~A.~C., Lefever, R.~R., Poniatowski, L.~G., et al.} 2023, 
\textit{\aap}, 670, A83

\bibitem[Sander et al.(2024)]{Sander+2024} 
{Sander, A.~A.~C., Bouret, J.-C., Bernini-Peron, M., et al.} 2024, 
\textit{\aap}, 689, A30

\bibitem[Villase{\~n}or et al.(2023)]{Villasenor+2023} 
{Villase{\~n}or, J.~I., Lennon, D.~J., Picco, A., et al.} 2023, 
\textit{\mnras}, 525, 5121

\bibitem[Vink et al.(2001)]{Vink+2001} 
{Vink, J.~S., de Koter, A., \& Lamers, H.~J.~G.~L.~M.} 2001, 
\textit{\aap}, 369, 574

\bibitem[Vink \& de Koter(2005)]{VinkdeKoter2005} 
{Vink, J.~S. \& de Koter, A.} 2005, 
\textit{\aap}, 442, 587

\bibitem[Vink et al.(2011)]{Vink+2011} 
{Vink, J.~S., Muijres, L.~E., Anthonisse, B., et al.} 2011, 
\textit{\aap}, 531, A132

\bibitem[Vink(2017)]{Vink2017} 
{Vink, J.~S.} 2017, 
\textit{\aap}, 607, L8

\bibitem[Vink \& Sander(2021)]{VinkSander2021} 
{Vink, J.~S. \& Sander, A.~A.~C.} 2021, 
\textit{\mnras}, 504, 2051

\bibitem[Vink et al.(2023)]{Vink+2023} 
{Vink, J.~S., Mehner, A., Crowther, P.~A., et al.} 2023, 
\textit{\aap}, 675, A154


\end{thebibliography}
\end{document}